\documentstyle[aps,preprint,epsfig,floats]{revtex}

\tightenlines

\begin{document}

\title{Comment to `The dependence of the anomalous $J/\psi$
suppression on the number of participant nucleons'}

\author{
A.P. Kostyuk$^{a,b}$,
%\footnote{E--mail: goren@th.physik.uni-frankfurt.de},
H. St\"ocker$^{a}$,
%\footnote{E--mail: stoecker@th.physik.uni-frankfurt.de}
and
W. Greiner$^{a}$
%\footnote{E--mail: greiner@th.physik.uni-frankfurt.de}
}

\address
{$^a$ Institut f\"ur Theoretische Physik, Universit\"at  Frankfurt,
Germany}

\address{$^b$ Bogolyubov Institute for Theoretical Physics,
Kyiv, Ukraine}

\date{\today}
\maketitle

\begin{abstract}
The recently published experimental dependence
of the $J/\psi$ suppression pattern in Pb+Pb collisions at the CERN SPS 
on the energy of zero degree calorimeter $E_{ZDC}$ are analyzed.
It is found that the data obtained within the `minimum bias' analysis
(using `theoretical Drell-Yan')
are at variance with the previously published 
experimental dependence of the same quantity on the 
transversal energy of neutral hadrons $E_T$. The discrepancy is related
to the moderate centrality region: $100 \alt N_p \alt 200$ ($N_p$ is the 
number of nucleon participants). This could result from  systematic 
experimental errors in the minimum bias sample.
A possible source of the errors may be contamination of the minimum bias sample 
by off-target interactions.
The data obtained within the standard analysis (using measured Drell-Yan
multiplicity) are found to be much 
less sensitive to the contamination.
\end{abstract}


Recently, the NA50 collaboration published new data on the
centrality dependence of the $J/\psi$ suppression pattern in Pb+Pb
collisions at CERN SPS \cite{NA50}. The centrality of the
collisions was estimated by measuring the energy of projectile
spectator nucleons $E_{ZDC}$ by a zero degree calorimeter. The purpose
of the present comment is to compare the new data with the ones
published by the NA50 collaboration previously
\cite{anomalous,threshold,evidence}, where the transverse energy
of produced neutral hadrons $E_T$ was used as the centrality
estimator. It is shown that the two sets of data (for brevity, 
we shall call them `$E_{ZDC}$-data' and `$E_T$-data', respectively),
having very similar qualitative behavior, seem, nevertheless, to be 
at variance on the quantitative level. This may point out a presence of
systematic errors in one of the data sets.

In each data set, the $J/\psi$ suppression pattern were obtained
in two different ways: by the standard analysis and the minimum
bias analysis. In the standard analysis, the ratio $R$ of $J/\psi$
multiplicity to the {\it physical} multiplicity of Drell-Yan pairs
was calculated from the measured dimuon spectra. In the minimum
bias analysis, the ratio $R$ was obtained by dividing the
experimentally measured ratio of the number of $J/\psi$ events to
the number of minimum bias events by the ´{\it theoretical} Drell-Yan'
given by
\begin{equation}\label{TheoDY}
\langle DY(E_T) \rangle = \frac{ \int_{0}^{\infty} d^2 b P(E_T|b)
\langle DY(b) \rangle P_{int}(b) }{\int_{0}^{\infty} d^2 b P(E_T|b)
 P_{int}(b)}
\end{equation}
Here $\langle DY(b) \rangle$ is the average number of Drell-Yan
dimuon pairs per Pb+Pb collision at impact parameter $b$.
$P_{int}(b)$ is the probability that an interaction between two
nuclei at impact parameter $b$ takes place. Both quantities were
calculated in the Glauber approach.

The $E_T$-distribution of events at fixed impact parameter $b$,
$P(E_T|b)$, was assumed to have the Gaussian form with the central
value and dispersion given by
\begin{equation}
\langle E_T(b) \rangle = q N_p(b) \ \ \ \ \ \
\sigma_{E_T}(b) = q \sqrt{a N_p(b)}
\end{equation}
Here $N_p(b)$ is the average number of nucleon participants in a Pb+Pb 
collision at impact factor $b$ calculated in the Glauber approach. The
parameter values $q=0.274$ GeV and $a=1.27$ \cite{Chaurand} were
fixed from the minimum bias transverse energy distribution
\cite{evidence}.

The $E_{ZDC}$-dependence of `theoretical Drell-Yan' was found from
the formula similar to (\ref{TheoDY}). The $E_{ZDC}$-distribution
$P(E_{ZDC}|b)$ was assumed to be a Gaussian with the central value
and dispersion given by Eqs. (1) and (2) of Ref. \cite{NA50}.

It is obvious that in absence of essential systematic errors, the
$E_T$- and $E_{ZDC}$-data should be consistent with each other. If
there exists a parameterization of the $b$-dependence of the
$J/\psi$ multiplicity $\langle J/\psi (b) \rangle$, such that the
$E_T$ dependence of $R$ computed from
\begin{equation}
R(E_T)=\frac{\int_{0}^{\infty} d^2 b P(E_T|b) \langle J/\psi (b)
\rangle P_{int}(b)}{\langle DY \rangle_{E_T}}
\end{equation}
agrees with the $E_T$-data, then the dependence of $R$ on
$E_{ZDC}$
\begin{equation}
R(E_{ZDC})=\frac{\int_{0}^{\infty} d^2 b P(E_{ZDC}|b) \langle J/\psi
(b) \rangle P_{int}(b)}{\langle DY \rangle_{E_{ZDC}}}
\end{equation}
should agree with the $E_{ZDC}$-data without any additional
parameter fit. And vice versa: a model fitted to the $E_{ZDC}$-data should 
automatically agree with the $E_T$-data.

To check, whether the two sets of the NA50 data indeed possess the
above property, we used three different models to parameterize
$\langle J/\psi (b) \rangle$. Two of them, the models A and B (see
below), were fitted to the $E_T$-data presented in Fig. 4 of 
Ref.\cite{evidence} and then were compared with the `minimum bias'
$E_{ZDC}$-data from Fig.5 of Ref.\cite{NA50}. The third one, the  
model C, was fitted to the `minimum bias'
$E_{ZDC}$-data and compared with the $E_T$-data.

{\bf Model A} is essentially the statistical coalescence model
(SCM) \cite{Br1,Go:00}, which assumes that charmonia are created at 
the final stage of the heavy ion reaction
from the charmed quark-antiquark pairs produced in hard
parton collisions at the initial stage. The free parameters of the SCM (see
Ref. \cite{Ko:01} for details) are $\sigma^{NN}_{c\bar{c}}$, the
in-medium modified \cite{hf_enh} cross section of open charm production in
nucleon-nucleon collisions, 
and $\eta$, the fraction of dimuon pairs,
originating from $J/\psi$ decays, that satisfies the kinematical
conditions of the NA50 spectrometer.

The SCM is known to be applicable only for sufficiently large
systems ($N_p > 100$). To parameterize the data in the low $E_T$
region, we assumed a smooth transition between the normal nuclear
suppression (NNS) to the statistical coalescence model:
\begin{equation}
\langle J/\psi (b) \rangle= \exp \left [- \zeta N_p^2(b) \right]
\langle J/\psi (b) \rangle_{NNS} + \left \{ 1 -  \exp \left [-
\zeta N_p^2(b) \right] \right \}  \langle J/\psi (b)
\rangle_{SCM}.
\end{equation}
This introduces an additional free parameter $\zeta$. The value of
NNS cross section was fixed at $\sigma_{abs}=6.4$ mb.

The drop of $R(E_T)$ at $E_T \agt 100$ GeV appears in SCM due to
fluctuations of the thermodynamical parameters and due to $E_T$
losses in the dimuon event sample relative to the minimum bias one
\cite{Ko:02}.

A good fit of the $E_T$-data ($\chi^2/\mbox{dof}=1.10$) is
obtained with
\begin{equation}
\begin{array}{lclcl}
\sigma^{NN}_{c\bar{c}}=3.67 \ \mu\mbox{b}  &\ \ \ &  \eta=0.129 &\ \ \ &
\zeta=1.04 \cdot 10^{-4}
\end{array}.
\end{equation}

{\bf Model B} is the geometrical model of charmonium suppression
\cite{JP,Mai}. It is assumed that all charmonia suffer the normal
nuclear suppression with the cross section $\sigma_{abs}$. Then all
excited charmonium states are destroyed in the region, where the
nucleon participant density in the plane transverse to the
collision axis exceeds the value $n_1$. This suppresses the
$J/\psi$ yield by a factor $X<1$, because excited charmonia
contribute to the final $J/\psi$ multiplicity through their decays. 
No charmonium state survives in the region, where the participant 
density exceeds the value $n_2$. Fluctuations of $E_T$
result in additional suppression at $E_T \agt 100$ GeV \cite{Mai}.
$E_T$ losses are ignored.

To make our parameterization as good  as possible we treated not only
$n_1$ and $n_2$, but also $\sigma_{abs}$ and $X$ as free
parameters. A very good fit ($\chi^2/\mbox{dof}=0.95$) is obtained
at
\begin{equation}
\begin{array}{lcl}
n_1 = 2.97 \mbox{ fm}^{-2}  &\ \ \ & n_2= 4.05 \mbox{ fm}^{-2} \\
\sigma_{abs}= 5.16 \mbox{ mb} &\ \ \ &  X= 0.366           
\end{array}
\end{equation}

\begin{figure}[p]
\begin{center}
%\vfill
%\leavevmode
\epsfig{file=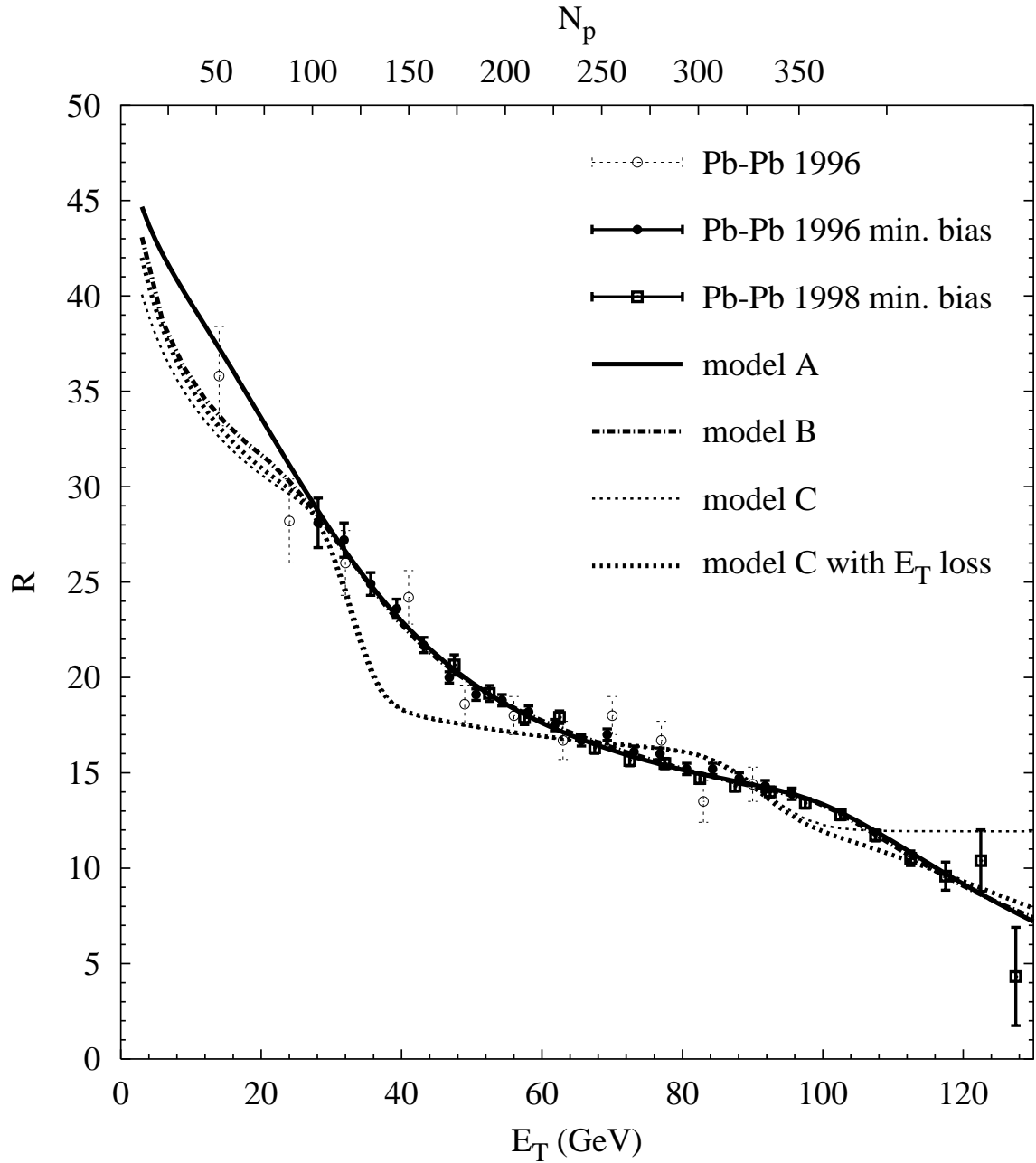,height=17cm} \mbox{}\\ \vfill
\mbox{}\\[1cm] \caption{The dependence of the $J/\psi$ to
Drell-Yan ratio $R$ on the transverse energy $E_T$. The
corresponding average number of participant nucleons $N_p$ are
shown on the upper axis. The points with error bars are the NA50
data. The lines correspond to different parametrizations. Models A
an B are fitted to the $E_T$ data set. Model C being fitted to the
$E_{ZDC}$ data set (see Fig. \ref{f2}) demonstrates disagreement
with the $E_T$ data. \label{f1}}
\end{center}
\end{figure}

In spite of the fact that the underlying ideas of the models A and
B are very different, the fits are completely consistent with
each other everywhere, except the low $E_T$ region, where only two
points with large statistical errors are available (see Fig.
\ref{f1}).

\begin{figure}[p]
\begin{center}
%\vfill
%\leavevmode
\epsfig{file=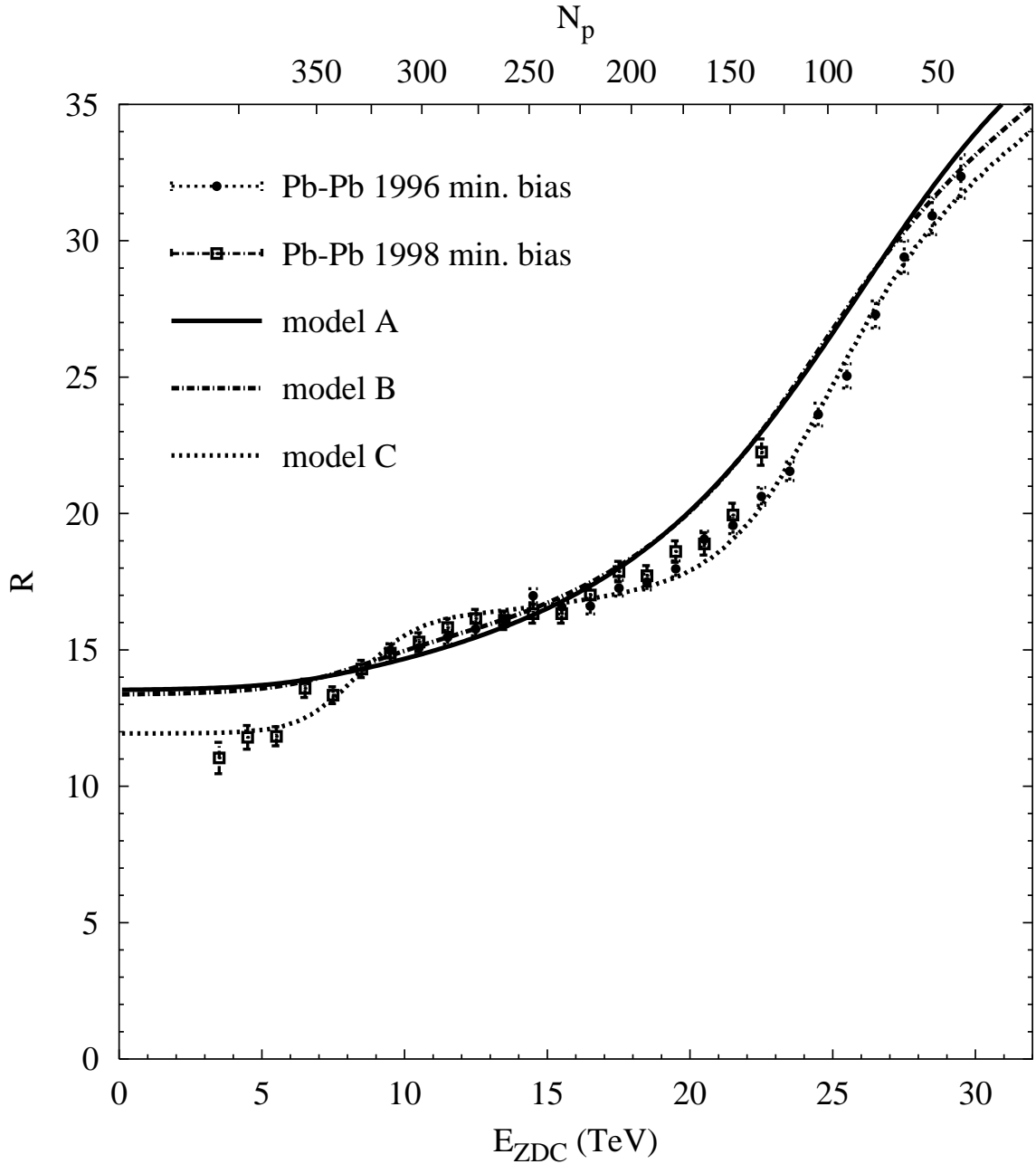,height=17cm}
\mbox{}\\
\vfill
\mbox{}\\[1cm]
\caption{The dependence of the $J/\psi$ to Drell-Yan
ratio $R$ on the energy of zero degree calorimeter $E_{ZDC}$.
The corresponding average
number of participant nucleons $N_p$ are shown on the upper axis.
The points with error bars are the NA50 `minimum bias' analysis data.
The lines correspond to different parameterizations. Model C is
fitted to the $E_{ZDC}$ data set.
Models A and B being fitted to the $E_T$-data (see Fig. \ref{f1})
demonstrate disagreement with the $E_{ZDC}$ data.\label{f2}}
\end{center}
\end{figure}

The both parameterizations are, however, in obvious disagreement with
the `minimum bias' $E_{ZDC}$-data: 
$\chi^2/\mbox{dof}=12.1$ and
$\chi^2/\mbox{dof}=11.5$ for the models A and B, respectively. The
strongest discrepancy is observed in the domain $100 \alt N_p \alt
200$ (see Fig. \ref{f2}).

It must be noted that the effects that were responsible for the
drop of the ratio R at $E_T \agt 100$ GeV, the fluctuations and
$E_T$ losses, do not influence the $E_{ZDC}$ dependence. Therefore,
neither of the models is able to reproduce the drop of $R(E_{ZDC})$
at $E_{ZDC} \alt 9$ TeV.

Let us go the opposite way, i.e. we 
fit the $E_{ZDC}$ data and check, whether the fit agrees with
the $E_T$ data. For this purpose, the model C will be
used.

{\bf Model C} was proposed by NA50 collaboration in
Ref.\cite{NA50}. The idea is similar to the model B, but the
parameter that controls the charmonium suppression is not the
nucleon participant density, but rather the number of
participants. It is assumed that at $N_p > N_{p1}$, the $J/\psi$
yield is suppressed by the factor of $X_1<1$ relative to the
normal nuclear suppression value. At $N_p > N_{p2}$ the remaining
$J/\psi$ yield is additionally suppressed by the factor $X_2$. A
minimum of $\chi^2$ is reached at
\begin{equation}
\begin{array}{lclcl}
N_{p1}=122  &\ \ \ & N_{p2}=335 &\ \ \ & \sigma_{abs}=5.47   \mbox{mb} \\  
X_1=0.670 &\ \ \ & X_2=0.752
\end{array}
\end{equation}
Although the quality of the fit is not very good
($\chi^2/\mbox{dof}=2.73$), the model satisfactory reproduces the shape
of the $E_{ZDC}$ data, including the drop 
of the ratio $R$ at $E_{ZDC} \alt 9$ TeV.

Comparing the model C with the $E_T$ data reveals strong
disagreement: $\chi^2/\mbox{dof}=13.3$. The strongest discrepancy is
again related to the region $100 \alt N_p \alt 200$ (see Fig.
\ref{f1}).  At large $E_T$,
although the drop of $R$ at $E_T \agt 100$ GeV is present on
the curve, it does not reproduce the behavior of the data. Account
for $E_T$ losses in the dimuon sample (similar to \cite{E_T_loss,Ko:02})
somewhat improves agreement
with the data, but still the shape of the drop is not reproduced.

The above consideration demonstrates that the two sets of
the NA50 data on the centrality dependence of the $J/\psi$
suppression pattern: the $E_T$-dependence in Fig. 4 of
Ref.\cite{evidence} and the $E_{ZDC}$-dependence in Fig. 5 of
Ref.\cite{NA50} are at variance with each other. This could mean that
at least one of the two sets may contain sizable systematic
errors.

In our opinion, the $E_{ZDC}$-data obtained within the minimum 
bias analysis are very likely to contain essential systematic 
errors because of possible contamination of the minimum
bias sample by off-target interactions. In the data analysis procedure
of Ref.\cite{NA50}, empty target runs were used to remove the 
contribution of the off-target interactions. This approach, however, 
does not take into account possible influence of target on the shape of the 
contamination. In fact, spectator fragments from an off-target collision 
may interact with the target. On the other hand, the spectator fragment from
a normal collision with a target nuclei may experience off-target interactions
before they are registered by the zero degree calorimeter. Therefore, 
in presence of the target, the off-target contribution 
may be shifted downwards in $E_{ZDC}$.

Indeed, the minimum bias
$E_T$ distribution perfectly agrees with the Glauber model (see
Fig. 1  of Ref.\cite{evidence}).\footnote{ 
Disagreement is observed
only in the low $E_T$ region, because the efficiency of the target
algorithm is less than unity there.} 
In contrast, the
agreement of the $E_{ZDC}$ minimum bias distribution seems to be not so 
perfect.
Unfortunately, the authors of Ref.\cite{NA50} do not quote the value of
$\chi^2/\mbox{dof}$, which characterizes the quality of the fit of 
the minimum bias $E_{ZDC}$ distribution (shown in Fig.1 of Ref.\cite{NA50}). 
Nevertheless, it is seen from the plot, that the experimental points 
lie slightly above the theoretical curves in the region 
$25 \agt E_{ZDC} \agt 18$ TeV (which corresponds
to $100 \alt N_p \alt 200$). Although the difference does not
exceeds even the size of the point symbols, it, nevertheless, indicates a 
sizable discrepancy, because of the logarithmic scale of the vertical axis of
the plot. Note, that at low $E_{ZDC}$, i.e. in the region of the `second
drop' of the $E_{ZDC}$-data, the theoretical models are also
in disagreement with the data. Therefore, the discrepancy of the two
sets of data in the region $100 \alt N_p \alt 200$ and the drop of
the ratio $R(E_{ZDC})$ at $E_{ZDC} \alt 9$ TeV may have a common origin:
contamination of the minimum bias sample by off-target
interactions.

\begin{figure}[p]
\begin{center}
%\vfill
%\leavevmode
\epsfig{file=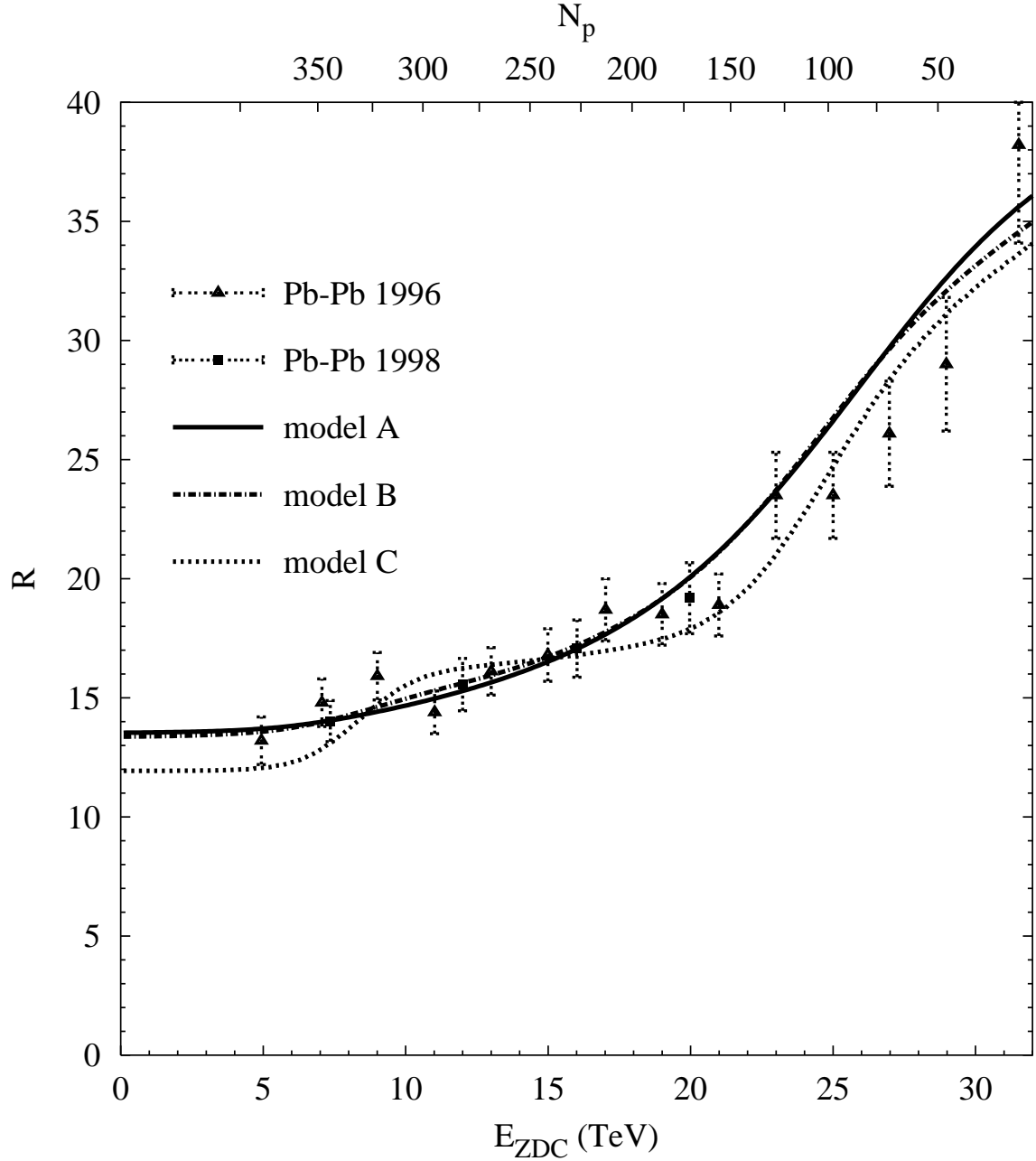,height=17cm}
\mbox{}\\
\vfill
\mbox{}\\[1cm]
\caption{The dependence of the $J/\psi$ to Drell-Yan
ratio $R$ on the energy of zero degree calorimeter $E_{ZDC}$.
The corresponding average
number of participant nucleons $N_p$ are shown on the upper axis.
The points with error bars are the NA50 standard analysis data.
The lines correspond to different parametrizations. 
Models A and B are fitted to the $E_T$-data (see Fig. \ref{f1})
Model C is fitted to the `minimum bias' $E_{ZDC}$ data (see Fig. \ref{f2}).
Due to rather large statistical errors, all three models may be considered 
to be consistent with the standard analysis data.\label{f3}}
\end{center}
\end{figure}

The standard analysis data appear to be less sensitive to the off-target
contamination. In fact, the models A and B are consistent with the standard
analysis $E_{ZDC}$-data (presented in Fig. 3 of \cite{NA50}): 
$\chi^2/\mbox{dof}=0.85$ and 
$\chi^2/\mbox{dof}=0.83$, respectively
(see Fig. \ref{f3}). No definite conclusion
is possible because of the large statistical errors. The model C may be 
also considered to be consistent with the standard analysis
data with slightly worse but still quite satisfactory fit quality:
$\chi^2/\mbox{dof}=1.10$. Nevertheless, no contradiction between $E_T$ and 
$E_{ZDC}$ is seen in the recently presented by the NA50 collaboration
standard analysis data from 
the year 2000 run, which have essentially smaller error bars \cite{Ramello}.
This allows to conclude that the minimum bias sample seem to be more 
sensitive to the off-target contaminations than the dimuon one.

The discrepancy between $E_T$ and $E_{ZDC}$ data are unlikely to be
related to  interactions of Pb nuclei and their fragments with air.
In fact, comparing of the data collected in the year 
2000 with the target in vacuum \cite{Ramello} with the data from
previous runs 
\cite{NA50,anomalous,threshold,evidence} indicates that interactions
with air influence both $E_T$ and $E_{ZDC}$ data and that the influence 
is related mostly to the peripheral domain $N_p \alt 100$.

The contaminations at moderate centrality, $100 \alt N_p \alt 200$, 
may come from interactions of reactions fragments with nuclei in zero
degree calorimeter. If such an interaction takes place, a fraction of 
the produced particles (those having large transversal momenta)
may leave the calorimeter without deposing their energy in it. 
This could cause 
sizable losses of zero-degree energy, which result in distortion
of the shape of the minimum bias sample. Therefore, the mentioned problem
will likely persist even in the new vacuum data, if they are analyzed
within the `minimum bias' procedure, unless losses of the zero degree energy
are taken into account.

We conclude, that the new data on the $E_{ZDC}$ dependence of 
the $J/\psi$ suppression pattern in Pb+Pb at the CERN SPS \cite{NA50}
are at variance with the other data published by the same collaboration 
\cite{anomalous,threshold,evidence}. This problem is related only to 
the data obtained within the `minimum bias' analysis.
A possible source of the discrepancy
might be distortion of the minimum bias sample most likely by
losses of zero degree energy because of 
interactions of reaction fragments with nuclei inside the zero degree 
calorimeter. 

\acknowledgments
We are grateful to R.~Arnaldi, P.~Bordalo, M.~Gonin, L.~Kluberg, 
M.~Mac Cormick, E.~Scomparin, N. Topilskaya for their interest to the paper and
for interesting discussions. 
We acknowledge the financial support of
the Alexander von Humboldt Foundation, DFG, GSI and BMBF, Germany.

\widetext

\nopagebreak

\end{document}